\documentclass[usenatbib]{mn2e}                  

\usepackage{graphicx}
\usepackage{amssymb}

\usepackage{aas_macros}

\title[Fermi acceleration and magnetic reconnection]{First order Fermi acceleration driven by magnetic reconnection}

\author[L. O'C. Drury]{L. O'C. Drury\\
Dublin Institute for Advanced Studies\\
School of Cosmic Physics\\
31 Fitzwilliam Place\\
Dublin 2\\
Ireland}

\begin{document}
\maketitle

\begin{abstract}
A box model is used to study first order Fermi acceleration driven by magnetic reconnection.  It is shown, at least in this simple model,  that the spectral index of the accelerated particles is related to the total compression in the same way as in diffusive shock acceleration and is not, as has been suggested, a universal $E^{-5/2}$ spectrum.  The acceleration time-scale is  estimated and some comments made about the applicability of the process.
\end{abstract}

\section{Introduction}
It has been suggested that the convergent flows associated with magnetic reconnection may drive a process of first order Fermi acceleration.  In particular, in \citet{2005A&A...441..845D} it is argued that such a process can occur and naturally produce an $E^{-5/2}$ power-law differential energy spectrum of accelerated particles.  More recently \citet{2010MNRAS.408L..46G} has invoked this process to explain the acceleration of the UHECRs (cosmic rays with $E>10^{19}\,\rm eV$). It has also been considered in \citet{2011arXiv1110.0557C} and \citet{2011P&SS...59..537L}.  It is thus of interest to examine this idea in more detail.

\section{The model of de Gouveia Dal Pino and Lazarian}

The basic idea is quite simple.  The reconnection layer is approximated as a thin planar sheet with plasma flowing in from both sides at the reconnection velocity $V_{\rm rec}$.  Charged particles with gyro-radii much larger than the thickness of the sheet are then assumed to freely diffuse in the inflowing plasma and, much as in the standard theory of diffusive shock acceleration, to gain energy every time they cross the reconnection layer.  As in the theory of shock acceleration, this relies on using a mixed system of coordinates where particle momenta (and associated energies) are measured with respect to the local co-moving plasma frame.  This has the advantage that all magneto-static scattering effects in the plasma on either side of the layer, responsible for isotropising the particle distributions and producing spatial diffusion, are then energy-neutral and the energy changes are entirely due to the frame-change on crossing.  It should be noted that as long as the distributions are close to isotropy and the sheet is assumed to be thin, direct acceleration by the electric field in the reconnection layer as suggested by \cite{Speiser:1965lr} is unimportant except for a very small set of particles on Speiser orbits. 

An argument essentially identical to that given by \cite{1978MNRAS.182..147B} for the case of shock acceleration then shows that on each crossing of the reconnection sheet the particle momentum or energy increases, averaging over an isotropic distribution, by an amount
\begin{equation}
{\Delta p\over p} = {\Delta E\over E} = {4\over 3} {V_{\rm rec}\over c}
\end{equation}
or twice this for a cycle of two crossings (bringing the particle back to the side it started on).
The fractional gain in momentum ($\Delta p/p$)  resulting from the frame change on crossing the discontinuity is easily found to be $1+\mu\beta$ where $\beta = 2 V_{\rm rec}/c$ is the dimensionless
velocity jump across the discontinuity and $\mu = \cos\theta$ is the cosine of the angle at which the particle crosses with respect to the normal.  Averaging over an isotropic distribution,
$\left<\mu\right> = 2\int_0^1 \mu^2 d\mu = 2/3$ then gives the above result.

de Gouveia Dal Pino and Lazarian then argue that "the escape probability of a particle from the acceleration zone is similar to the one computed for a shock front" and that the probability (per cycle of crossings) of staying within reconnection region is 
\begin{equation}
P= 1- {4V_{\rm rec}\over c}.
\end{equation}
Now if the energy per cycle increases by a factor $1+\alpha$ and the probability of remaining in the system per cycle is $1-\beta$ it is easy to show \cite{1978MNRAS.182..147B} that the steady state differential energy spectrum is 
$N(E)\propto E^{-1-\beta/\alpha}$ from which they deduce a spectrum of $E^{-5/2}$.

There are two serious objections to this analysis. First, in calculating the energy gain they assume that the flow is entirely compressive and neglect the outflow from the reconnection region which must be associated with a divergence of the flow field (and associated energy losses).  Secondly, their assumption that the escape rate is the same as that from a shock is entirely without justification.  Even it it is accepted as an order of magnitude estimate, it certainly does not justify the precise value required to get the spectral index of $-5/2$.  Nevertheless the idea is interesting and worth pursuing.
\cite{2010MNRAS.408L..46G}
considers the ultra-relativistic version of the process, but focuses on the maximum attainable energy and does not consider the spectrum.

\section{A box model treatment}

\begin{figure}
\begin{center}
\includegraphics[width=\hsize]{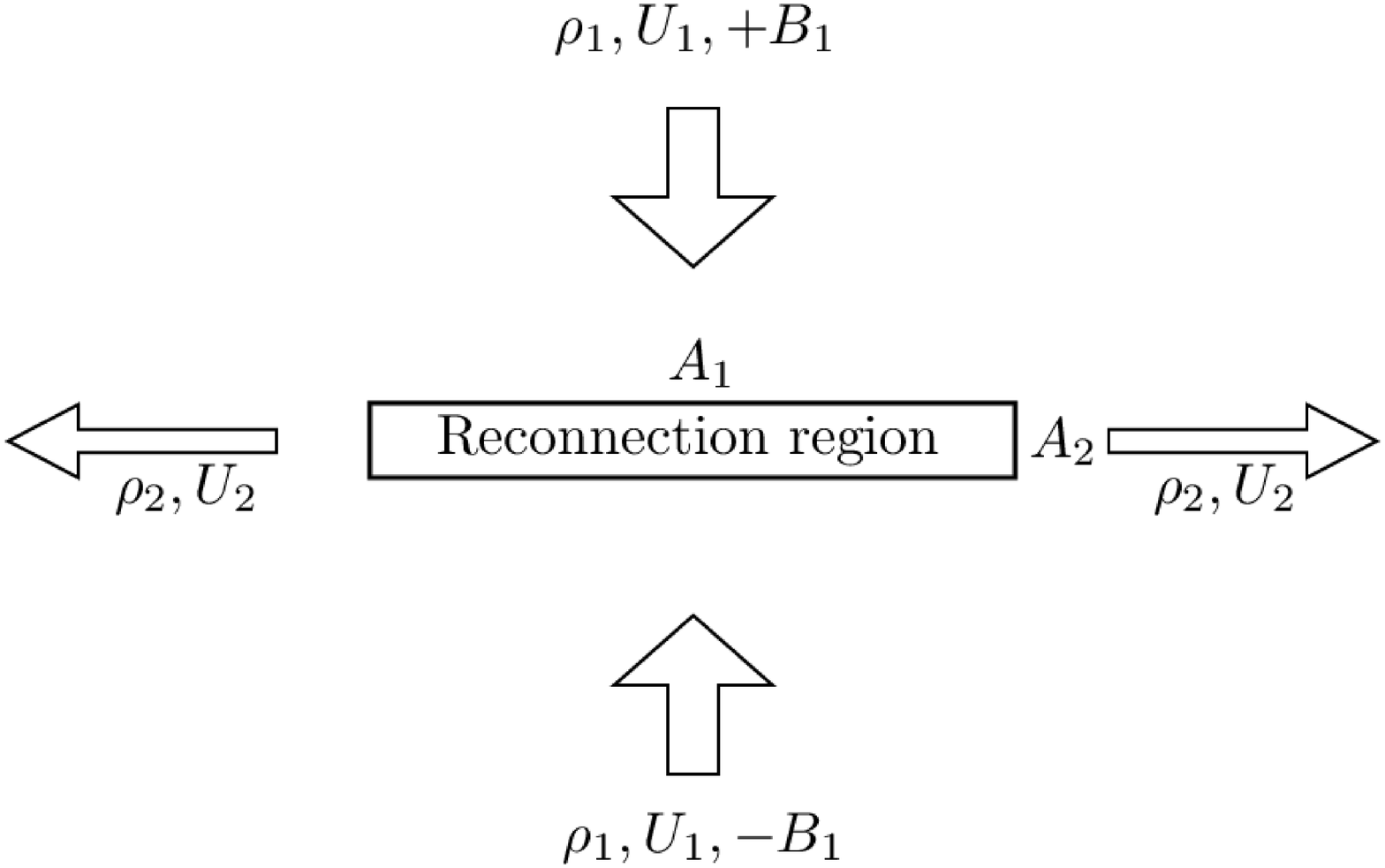}
\caption{A cartoon sketch of a generic magnetic reconnection system}
\label{Cartoon1}
\end{center}
\end{figure}

Let us model the reconnection process as a `black box' system into which highly magnetized plasma flows symmetrically from above and below and out of the sides of which emerge two `jets' of plasma with little internal magnetic field (see Figure \ref{Cartoon1}).

There are two conservation laws that can be applied to constrain this system.  First, the flux of mass in has to equal the flux out if the system is quasi-steady,
\begin{equation}
\rho_1 U_1 A_1 = \rho_2 U_2 A_2 = \dot M,
\end{equation}
where the incoming plasma has density $\rho_1$, velocity $U_1$ and the cross-sectional area of the reconnection region for the inflow is $A_1$ and similarly for the outflow.

Secondly, the energy flux into the system has to equal that out. 
Noting that the magnetic energy density is related to the local Alfven speed, $V_A$ by
\begin{equation}
{B^2\over 2\mu_0} = {1\over2}\rho V_A^2,
\end{equation}
and if we assume that magnetic energy density dominates in the inflow and kinetic energy density in the outflow, this gives
\begin{equation}
{1\over 2} \rho_1 V_A^2 U_1 A_1 = {1\over 2} \dot M V_A^2 \approx 
{1\over 2} \rho_2 U_2^3 A_2 = {1\over 2} \dot M U_2^2,
\end{equation}
from which the well know result follows that the outflow velocity is of order the Alfven speed in the inflow.

Let us now consider the acceleration of particles due to the convergence of the flow in the reconnection region (which we denote as $R$) following the method of the simplified `box model' of shock acceleration as discussed in \cite{1999A&A...347..370D}.  The upwards flux of particles in momentum is simply given by
\begin{equation}
\Phi = \int_R 4\pi p^2 f(p) {p\over 3} (-\nabla\cdot U)
\end{equation}
and if we assume that for sufficiently high-energy particles the particle distribution is essentially uniform within the reconnection region then we get
\begin{eqnarray}
\Phi &=& {4\pi p^3\over 3}  f(p)  \int_R(-\nabla\cdot U) \\
&=& - {4\pi p^3\over 3} f(p) \int_{\partial R} U\\
&=& {4\pi p^3\over 3} f(p) \left[2A_1 U_1 - 2 A_2 U_2\right]
\end{eqnarray}
As in the shock case the steady spectrum can now be found by equating the divergence of the upwards flux in momentum to the losses carried out by the outflow,
\begin{equation}
{\partial \Phi\over\partial p} = - 2 A_2 U_2 4\pi p^2 f(p)
\end{equation}
which, after a little algebra, gives the result
\begin{equation}
{\partial \ln f\over \partial \ln p} = - {3A_1 U_1\over A_1 U_1 - A_2 U_2}.
\end{equation}
In fact this is precisely the same result as in shock acceleration if we express it in terms of the compression in the system. Using the fact that
\begin{equation}
A_1 U_1 = \dot M/\rho_1, \qquad A_2 U_2 = \dot M/\rho_2
\end{equation}
one easily finds that
\begin{equation}
{\partial \ln f\over \partial \ln p} = - {3r\over r-1},\qquad r={\rho_2\over\rho_1}.
\end{equation}

The question is thus how compressive can the process of magnetic reconnection be?   If little energy is used to heat the plasma and the conversion is essentially one of magnetic energy to kinetic energy it is clear that it can be very compressive.  The only constraint appears to be that the outflow not be significantly over pressured relative to the environment, and for a strongly magnetized inflow this is a very weak constraint.   Thus in general one expects rather large values of the compression ratio, certainly larger than the four canonically assumed for shocks, and thus extremely hard spectra tending in the limit to
\begin{equation}
f(p) \propto p^{-3}
\end{equation}
or a differential energy spectrum of $N(E)\propto E^{-1}$.  Of course such hard spectra rapidly lead of energy divergences and non-linear effects, but at least at the test-particle level it would appear that reconnection regions may be efficient sites for producing hard spectra by Fermi acceleration. 

It is also quite straightforward to derive the acceleration time-scale.  If the accelerated particles penetrate a distance $L_1$ `upstream' into the inflow and $L_2$ `downstream' into the outflow then it is easy to show that the acceleration time scale is
\begin{equation}
t_{\rm acc} =  3 {A_1 L_1 + A_2 L_2 + W\over A_1 U_1 - A_2 U_2}
\end{equation}
where $W$ is the volume of the reconnection region;
this again is a straightforward generalization of the result for shock acceleration and physically just expresses the fact that the upward flux in momentum takes a certain amount of time to fill the acceleration volume.

The above analysis, as is the case also in most discussion of the test-particle theory of shock acceleration, glosses over the question of where the particles come from (the injection problem).  It simply assumes that there is some injection process operating at energies lower than those being discussed which produces a population of supra-thermal particles which can then be further accelerated by this Fermi process.  In the case of reconnection this could be plasma physical processes related to the micro-physics of the actual reconnection, or it could be a pre-existing population of ambient particles (ambient cosmic rays).  

\section{Conclusions}

This approximate analysis suggests that reconnection can indeed drive a process of Fermi acceleration in close analogy to diffusive shock acceleration \citep{Krymskii:1977lr,Blandford:1978lr,Axford:1977qy,1978MNRAS.182..147B} even to the extent of having the same relation between total compression and the spectral index of the accelerated particles (at least to a first approximation).  Because there is no constraint from momentum conservation (the total momentum being identically zero by symmetry) the total compression will normally be quite high leading to rather hard spectra.  The suggestion by \cite{2005A&A...441..845D} of a universal $E^{-5/2}$ spectrum is not supported by this analysis.

The acceleration time scale is similar to that for shock acceleration, and the constraints on maximum energy due to the finite age and size of the reconnection region will thus be comparable to those in shock acceleration.  In general the much smaller scale of reconnection regions compared to shock structures will thus lead to significantly lower maximum energies (the dominant limiting loss process being diffusion out of the sides of the reconnection region).   Strictly the model developed here requires that the particle diffusion length scale in the inflow be large compared to the thickness of the reconnection sheet, but small compared to its lateral extent; thus it may only apply in a rather restricted energy range (this should be contrasted with the case of shock acceleration where the scale separation is much more extreme).  However the particles accelerated in the reconnection regions may be available as a seed population for further shock acceleration in many systems, and the fast time-scales and hard spectra may be of interest in models of flaring activity.

Finally it is interesting to note that for reconnection in an electron-positron pair plasma infinite compression is in theory possible by allowing the pairs to annihilate in the compressed reconnection sheet.  This may be relevant to acceleration by reconnection in pulsar winds.  In this case an extension of the model to include synchrotron losses would be required.


\bibliographystyle{mn2e.bst}

\bibliography{Reconnection}

\end{document}